\documentclass[%
 aip,
 amsmath,amssymb,
 reprint,%
]{revtex4-1}

\usepackage{graphicx}
\usepackage{dcolumn}
\usepackage{bm}

\usepackage[utf8]{inputenc}
\usepackage[T1]{fontenc}
\usepackage{mathptmx}

\usepackage{amsmath,amssymb,amsfonts}
\usepackage[export]{adjustbox}
\usepackage[format=plain,
      justification=RaggedRight,
      singlelinecheck=false]{caption}
\usepackage{subcaption}
\usepackage{textcomp}
\usepackage{xcolor}
\usepackage{amsthm}
\usepackage{enumitem}

    \newtheorem{lemma}{Lemma}
    
    \makeatletter
    \newcommand{\superimpose}[2]{{\ooalign{$#1\@firstoftwo#2$\cr\hfil$#1\@secondoftwo#2$\hfil\cr}}}
    \makeatother

\begin{document}

\preprint{AIP/123-QED}


\title{Statistical Analysis of Read-back Signals in Magnetic Recording on Granular Media}

\author{Florian Slanovc}
\email{florian.slanovc@univie.ac.at}
 \affiliation{Physics of Functional Materials, Faculty of Physics, University of Vienna, Austria.}
\author{Christoph Vogler}%
\email{christoph.vogler@univie.ac.at}
 \affiliation{Physics of Functional Materials, Faculty of Physics, University of Vienna, Austria.}
\author{Olivia Muthsam}
\email{olivia.muthsam@univie.ac.at}
 \affiliation{Physics of Functional Materials, Faculty of Physics, University of Vienna, Austria.}
\author{Dieter Suess}
\email{dieter.suess@univie.ac.at}
 \affiliation{Physics of Functional Materials, Faculty of Physics, University of Vienna, Austria.}

\date{\today}

\begin{abstract}

The comprehensive simulation of magnetic recording, including the write and read-back process, on granular media becomes computationally expensive if the magnetization dynamics of each grain are explicitly computed. In addition, in heat-assisted magnetic recording, the writing of a single track becomes a random process since the temperature must be considered and thermal noise is involved. Further, varying grain structures of various granular media must also be taken into account to obtain correct statistics for the final read-back signal. Hence, it requires many repetitions of the write process to investigate the mean signal as well as the noise.


This work presents a method that improves the statistical evaluation of the whole recording process. The idea is to avoid writing the magnetization to one of its binary states. Instead, we assign each grain its probability of occupying one of its stable states, which can be calculated in advance in terms of a switching probability phase diagram. In the read-back process, we combine the probabilities to calculate a mean signal and its variance. Afterwards, repetitions on different media lead to the final read-back signal.


Using a recording example, we show that the statistical behavior of the evaluated signal-to-noise ratio can be significantly improved by applying this probability mapping method, while the computational effort remains low.

\end{abstract}

\maketitle

\section{Introduction}
In state-of-the-art hard disk drives, which are used as storage media in personal computers as well as in server systems, the information is written as a sequence of bits on a granular medium consisting of magnetic grains. During writing, an applied magnetic field switches the magnetization direction of the magnetic grains in a specific direction\cite{b2}. Afterwards this direction can be detected by a reader module via the magnetic stray field. Due to the massive amount of data that is permanently produced, new technologies are required to increase the storage density of hard drives~\cite{b10,b5,b13,b9,b7}. Micromagnetic simulation of write and read-back processes is therefore a valuable tool for analyzing new ideas~\cite{b3}. Nevertheless, simulations of the entire write and read cycle are computationally expensive. Additionally, concepts such as heat-assisted magnetic recording (HAMR)\cite{b13} must consider temperature, which makes the underlying equation of the magnetization dynamics a stochastic partial differential equation that further increases the required computational effort~\cite{b11}. Due to the stochastic nature of the solution obtained, repeated simulations are required to reduce the statistical error of the observed results to a sufficient low level\cite{b1,b4,b6}. In this work we present an efficient calculation method of the signal-to-noise ratio (SNR) based on probability theory. Instead of writing concrete bit series on granular media, we allocate every grain a certain switching probability. In the read-back procedure, those probabilities are further processed to obtain the signal and noise value of the written bit sequence. The goal of this method is a significant improvement of the statistical error within the SNR evaluation of a given writing and read-back procedure. This reduces the number of necessary computation-intensive writing and read-back simulations required, resulting in an accurate SNR value.

\section{Signal-to-Noise Ratio}
Granular media for magnetic recording consist of single magnetic grains with strong uniaxial anisotropy perpendicular to the film plane, separated by non-magnetic grain boundaries. Due to the high uniaxial anisotropy of the grains, the assumption is valid that the magnetization of every single grain points either in positive or negative $z$-direction (perpendicular direction). Although each grain can only have two magnetic states $-1$ and $+1$, in reality writing a bit series has a stochastic nature. There are two reasons: First, in areas at the transitions between different bits, grains can have different states after consecutive writing processes, despite the fact that the same medium with the same grain pattern is used. Such grains have the probability to occupy the states $-1$ and $+1$ in the range of $[0,100]$\,\%. Additionally, the randomized positions of grains in different granular media leads to a further stochastic effect, because the probability of the magnetization direction of each grain depends on the position of the grain within the medium during the writing process. A read-back module measures the magnetization of the grains across the bit pattern and produces a corresponding read-back signal $V(x)$ as a function of the down-track position $x$. Since the magnetization of the individual grains is random, $V(x)$ is a random variable with certain expectation value $\mathbb E[V(x)]$ and variance $\mathbb V[V(x)]$. The signal power (SP) and the noise power (NP) of the whole bit pattern between the down-track positions $x_\text{start}$ and $x_\text{end}$ can be defined as

\begin{align}
\text{SP} = \int_{x_\text{start}}^{x_\text{end}}\mathbb E[V(x)^2]\ \text{d}x, \label{eq::sp} \\
\text{NP} = \int_{x_\text{start}}^{x_\text{end}} \mathbb V[V(x)]\ \text{d}x. \label{eq::np}
\end{align}

The quality criterium for a written bit series is the signal-to-noise ratio (SNR) defined by

\begin{align}
\text{SNR} = \frac{\text{SP}}{\text{NP}}.
\end{align}

The determination of SP and NP via measurement or simulation is an important goal of magnetic recording as in Refs.~\onlinecite{b1,b4}. In this paper we investigate and compare different numerical approaches based on statistic and probability calculation to determine these values in a computational accurate and non-expensive way. We further demonstrate the approaches on the example of simulating a writing process of heat-assisted magnetic recording, where the switching probability of grains is in particular affected by additional thermal fluctuation.

\section{Switching Probability Phase Diagram for Single Grains}\label{sec::switchingprobability}
\begin{figure}[htbp]
\includegraphics[width=0.45\textwidth]{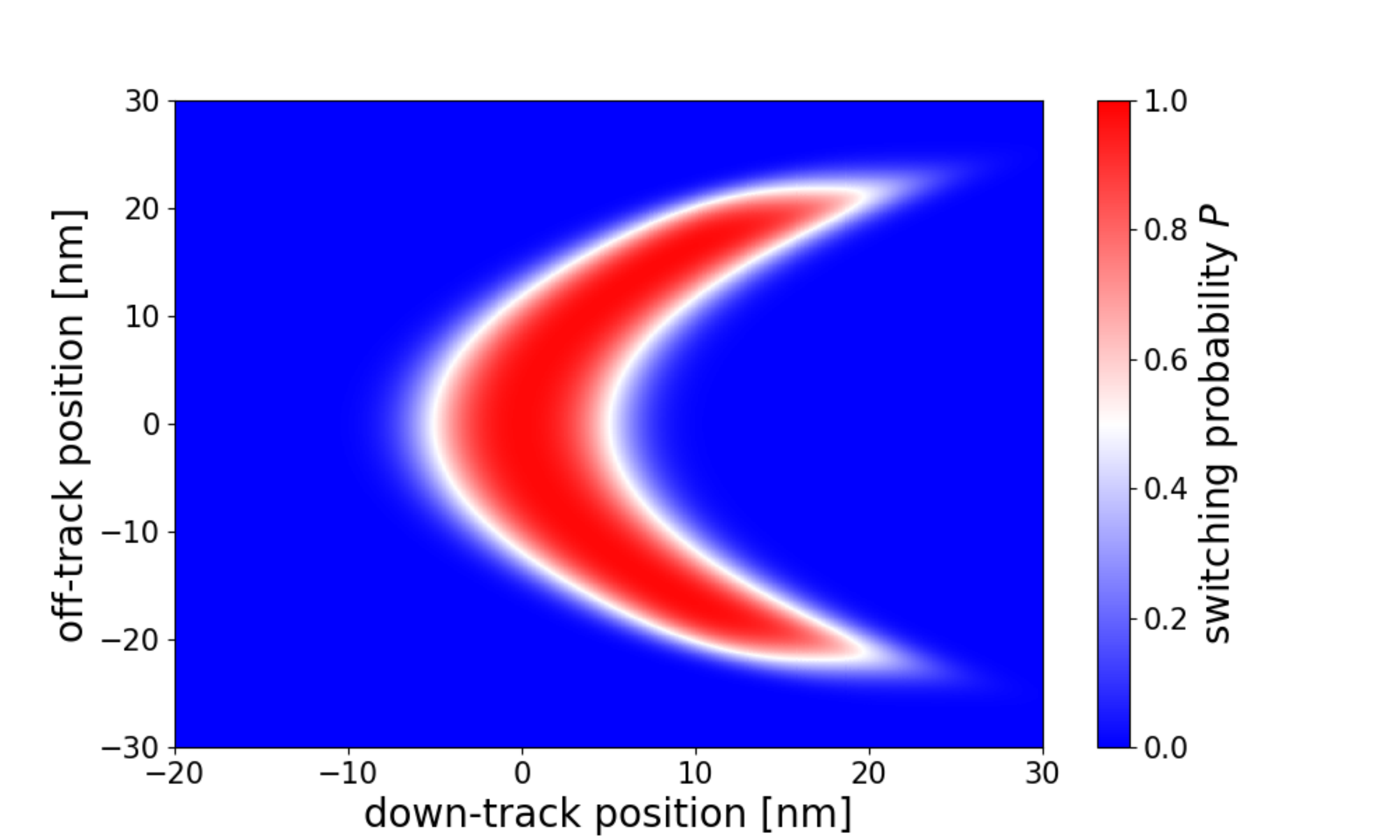}
\caption{Example for a switching probability phase diagram of one single grain.}
\label{fig::phasenplot1}
\end{figure}

The investigations in this work are based on the concept of switching probability phase diagrams as presented in Ref.~\onlinecite{b7}. The idea is the calculation of a single grain's switching probability dependent on its possible position on the recording track within the writing process. The advantage of this approach is that the phase diagram has to be computed only once and the switching probability can afterwards be extracted for an arbitrary amount of magnetic grains with no further computational effort. An example for the possible shape of such a diagram is given in Fig. \ref{fig::phasenplot1}. In the following, we always assume that such a phase diagram is available. Since a sequence of two or more successive down or up bits can be summed up to one writing operation with multiple bit length, we split the writing process of each bit pattern to $S$ sequential $-1$ or $+1$ mappings of the phase diagram onto the magnetic medium. The detailed realization of this sequential writing process depends on the different modeling approaches and will be discussed in the following sections.

\section{Magnetization Mapping on Grains}\label{sec::writing}

\subsection{Writing Process}\label{sec::writingprocess1}

For each bit pattern containing $G$ magnetic grains, we use the phase diagram to extract the switching probability $p_j^i$ for magnetic grain $i, (i=1,\ldots,G)$ in the $j$-th mapping step $(j=1,\ldots,S)$ of the bit pattern according to Sec. \ref{sec::switchingprobability}. Due to the assumption of only two possible magnetic states $m^i=-1$ and $m^i=+1$ of grain $i$, we assume randomized initialization of the grains, i.e.

\begin{align}
m^i_0 := \text{rand}^i_0(0.5),
\end{align}

where $\text{rand}(p)$ denotes the outcome of a random experiment with possible outputs $-1$ and $+1$ with probabilities $1-p$ and $p$. We then update the magnetization in each writing step $j$ recursively as it follows:\\
For given magnetization $m_{j-1}^i$ from the previous step, we calculate $m_{j}^i$ by differing two cases:
\begin{itemize}
\item If the $j$-th step writes a bit in $-1$ direction, we set:
\begin{itemize}
\item $m^i_j := -1$, if $m^i_{j-1} = -1$
\item $m^i_j := \text{rand}^i_j(1-p_j^i)$, if $m^i_{j-1} = +1$
\end{itemize}
\item If the $j$-th step writes a bit in $+1$ direction, we set:
\begin{itemize}
\item $m^i_j :=  \text{rand}^i_{j}(p_j^i)$, if $m^{i}_{j-1} = -1$
\item $m^i_j := +1$, if $m^i_{j-1} = +1$
\end{itemize}
\end{itemize}

\begin{figure}[htbp]
\centerline{\includegraphics[width=0.50\textwidth]{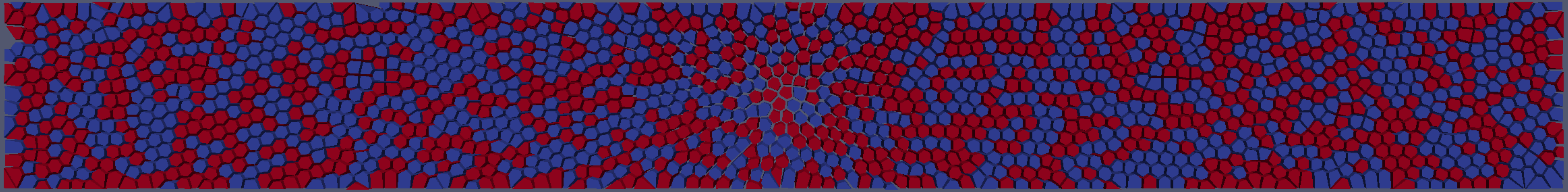}}
\centerline{\includegraphics[width=0.50\textwidth]{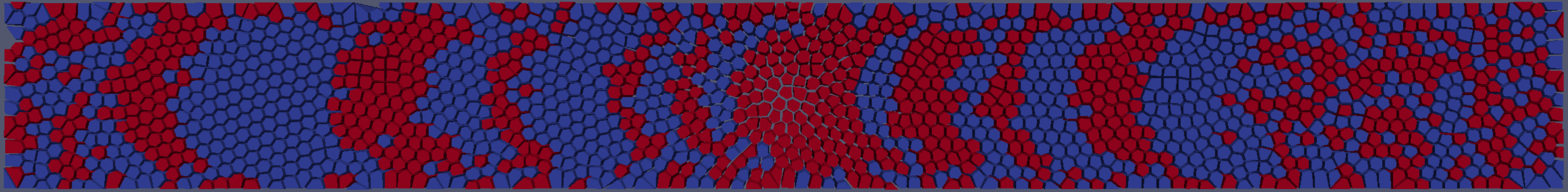}}
\caption{Top: Randomly initialized granular medium ($500\text{ nm}\times 60\text{ nm}$ and a thickness of $8$ nm) with grain diameter of $4$ nm and $1$ nm gap between neighboring grains. Bottom: Granular medium after the simulated writing process of a pseudo-random bit sequence according to Ref.~\onlinecite{b3}.}
\label{fig::writing}
\end{figure}

Doing $S$ mapping steps, this procedure leads to a sequence of grain magnetization and we receive the final magnetization of every grain $m^i := m^i_S\in\{-1,+1\}$. The result of such a writing procedure is shown in Fig. \ref{fig::writing}.

 \subsection{Read-back Process}
 
The reader module is defined via its sensitivity function, as defined in Ref.~\onlinecite{b8}. The voltage $V(x)$ of the reader in down-track position $x$ is given by the integral in Ref.~\onlinecite[Eq.~(1)]{b8}
\begin{align}
V(x) = c_1 \cdot \int \overrightarrow{H}\cdot \overrightarrow{M}\ \text{d}V_m
\end{align}
with the reader's sensitivity function $\overrightarrow{H}$, the magnetization of the media $\overrightarrow{M}$ and a reader dependent constant $c_1$. Since the magnetization is assumed to have negligible dependence on the $z$ direction and grains have strong uniaxial anisotropy, it degenerates to an area integral in the form
\begin{align}
V(x) = \widetilde{c}_1 \cdot \int H_z \cdot M_z \ \text{d}A_m
\end{align}
with a constant $\widetilde{c}_1$, which does not affect the SNR value, because both SP and NP are homogeneous of degree $2$ in $V$ and therefore the constant $\widetilde{c}_1$ cancels in the ratio $\text{SNR}=\text{SP}/\text{NP}$. Without loss of generality we set $\widetilde{c}_1:=1$. We further simplify the notation by skipping the $z$-index and set $H := H_z$ and $M := M_z$. The possible values for $M$ are $-1, +1$ or $0$ depending whether the position is located within a grain with $m^i = \pm 1$ or within the grain boundary. 
We can therefore rewrite the integral to a sum over the grains
\begin{align}\label{eq::signal2}
V(x) = \sum_{i=1}^G \underbrace{\left(\int_{\text{grain}_i} H\ \text{d}A_m \right)}_{=:H_G^i} \cdot m^i = \sum_{i=1}^G H_G^i \cdot m^i ,
\end{align}
where $H_G^i$ denotes the sensitivity across the grain $i$ for the reader module in position $x$. Moving the sensitivity function across the medium gives the detected read-back signal of the whole bit pattern.
 

 \subsection{Statistical Evaluation}\label{sec::statistic}
 
Repeating the previous step on $N$ granular media with different random grain structure and different outcomes of the random experiment $\text{rand}(p)$ (determining the magnetization according to the switching probabilities) leads to $N$ different signal trajectories $V_k$ $(k=1,\ldots, N)$. For large $N$ we can assume to receive a good estimation of SP and NP by substituting the squared expectation value and the variance in Eq. \eqref{eq::sp} and \eqref{eq::np} by their unbiased estimators

\begin{align}
\mathbb E[V(x)^2] &\approx \frac{1}{N}\overline{V(x)^2},\label{eq::erwartungswert}\\
\mathbb V[V(x)] &\approx \frac{1}{N-1} \sum_{k=1}^N \left( (V_k(x) - \overline{V(x)}\right)^2. \label{eq::varianz}
\end{align}

Due to the fact that we deal with two uncertainties, the grain magnetization according to the random experiment and the position of the grains in each granular medium, the amount of necessary repetitions $N$ might be relatively large to receive a good statistic for both.

\section{Probability Mapping on Grains}\label{sec::mapping}
\subsection{Writing Process}\label{sec::writingprocess2}
\begin{figure}[htbp]
 \begin{subfigure}{0.24\textwidth}
\centerline{\includegraphics[width=1.00\textwidth,frame]{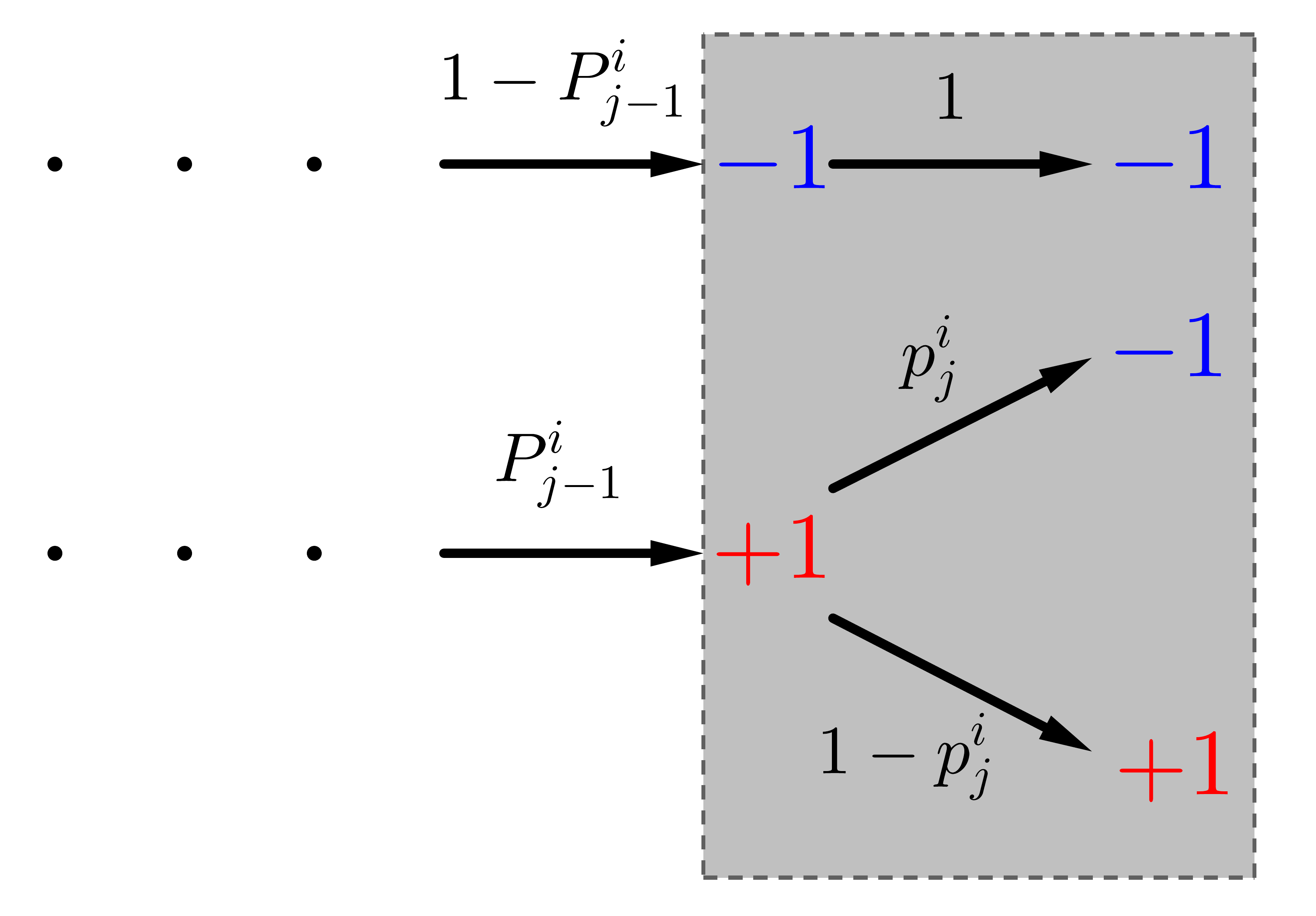}}
 \captionsetup{justification=centering}
\caption{}
\end{subfigure}~\begin{subfigure}{0.24\textwidth}
\centerline{\includegraphics[width=1.00\textwidth,frame]{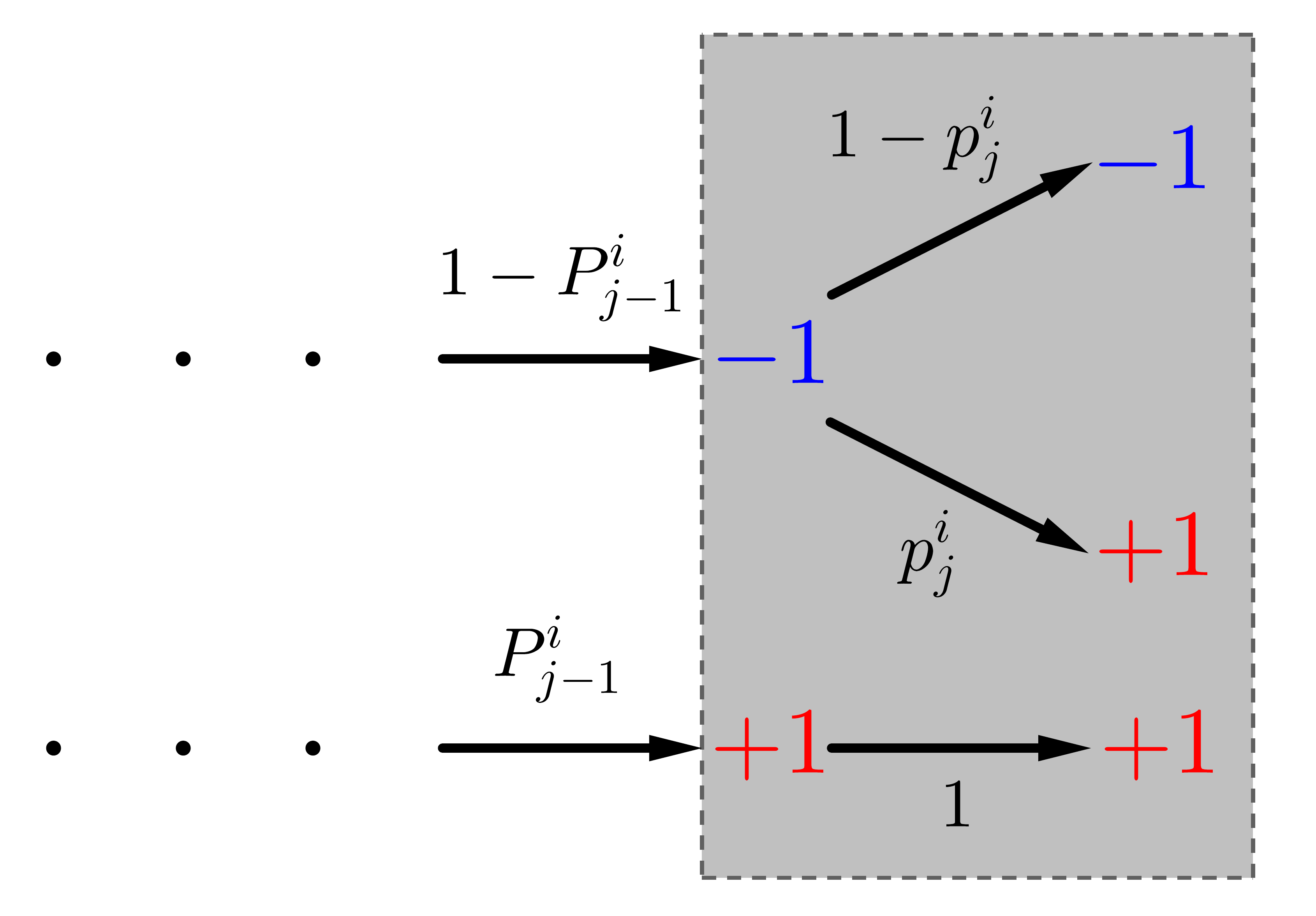}}
 \captionsetup{justification=centering}
\caption{}
\end{subfigure}
\caption{Tree diagram of the different decisions and their corresponding switching probabilities. The gray background marks the $j$-th writing step. (a): Writing process of a down-bit. (b): Writing process of an up-bit. \\ Based on to the previous magnetization of the grain with probabilities $1-P_{j-1}^i$ and $P_{j-1}^i$ respectively, the magnetization gets either changed from $-1$ to $+1$ or vice versa with switching probability $p_j^i$ or remains in its original direction with probability $1$.}
\label{fig::baum}
\end{figure}

In contrast to the previous magnetization mapping method, we now avoid setting the magnetic states to $-1$ or $+1$ in every writing step. Instead, we use probability analysis calculation laws to determine the overall probability for the $-1$ or $+1$ state after the entire writing process. This approach allows to calculate the expectation value and variance of the magnetization of each grain and therefore also for the whole read-back signal afterwards. As before, we consider the writing process as a sequence of $S$ independent mapping steps and in step $j$, the switching probability of the $i$-th grain is denoted by $p_j^i$ and again extracted from the phase diagram. We further denote the probability of grain $i$ for magnetization $+1$ after $j$ mapping steps with $P_j^i$. Since we assume only two possible states, the probability for magnetization $-1$ is therefore $1-P_j^i$. We initialize with random magnetization, i.e. $P_0^i := 0.5$ for all $i=1,\ldots,G$ and update the magnetization in each writing step $j$ recursively as it follows:\\
For given magnetization probability $P_{j-1}^i$, we calculate $P_{j}^i$ by differing two cases, which are also illustrated in Fig. \ref{fig::baum}. Due to our the definition of $P_{j}^i$, we aim in both cases for the probability of final state $+1$:
\begin{itemize}
\item If the $j$-th step writes a bit in $-1$ direction, we set $P_{j}^i := P_{j-1}^i \cdot (1-p_j^i)$ (see Fig. \ref{fig::baum} (a)).
\item If the $j$-th step writes a bit in $+1$ direction, we set $P_{j}^i := P_{j-1}^i + (1-P_{j-1}^i) \cdot p_j^i$ (see Fig. \ref{fig::baum} (b)).
\end{itemize}
After $S$ steps we receive the final magnetization probabilities of the whole writing process $P^i := P_S^i$. These probabilities allow to further calculate the expectation value and variance of the magnetization $m^i$ for each grain $i = 1,\ldots, G$ via the formulas
\begin{align}
\mathbb E[m^i] &= -1 \cdot (1-P^i) + 1 \cdot P^i = 2\cdot P^i - 1,\label{eq::e}\\
\begin{split}
\mathbb V[m^i] &= \mathbb E[(m^i)^2] - \mathbb E[m^i]^2 = 1 - (2\cdot P^i - 1)^2\\
& = 4\cdot P^i \cdot (1-P^i).
\end{split}\label{eq::var}
\end{align}
Note that in contrast to the previous section, $m^i$ is here regarded as a random variable and not determined by random numbers, so we have no statistical error so far.

 \subsection{Read-back Process}
The derivation of Eq.~\eqref{eq::signal2} again holds with the only difference that $m^i$ and therefore also $V(x)$ are random variables in this section. By applying $\mathbb E(\cdot)$ and $\mathbb V(\cdot)$ on both sides of the equation and using that $H_G^i$ is a deterministic value, we receive the expressions
\begin{align}
\mathbb E[V(x)] &= \sum_{i=1}^G H_G^i \cdot \mathbb E[m^i],\label{eq::1stmoment} \\
\mathbb V[V(x)] &= \sum_{i=1}^G \left(H_G^i\right)^2 \cdot \mathbb V[m^i].\label{eq::variance}
\end{align}
Additional, we can also calculate the second moment of $V(x)$ by the well known identity
\begin{align}
\mathbb E[V(x)^2] &= \mathbb V[V(x)] + \mathbb E[V(x)]^2. \label{eq::2ndmoment}
\end{align}

 \subsection{Statistical Evaluation}\label{sec::statistic2}
 We again repeat the writing and read-back process on $N$ different granular media. The outcome are $N$ different signal trajectories $V_k$ $(k=1,\ldots, N)$, where $V_k(x)$ denotes a random variable for every $k$ and down-track position $x$. The results of the previous subsection hold for each single medium. To combine them to a final SNR outcome, we state the following lemma for a discrete random variable in a specified random experiment:
\begin{lemma}\label{lem::momente}
Let $X_k$ be a discrete random variable for all $k=1,\ldots,N$ with corresponding probability distributions $P(X_k = x_{k\ell}) = p_{k\ell}$ for all $\ell = 1,\ldots,n_k$. We consider the two-step random experiment: 
\begin{itemize}
\item Choose $X_k \in \{X_1,\ldots,X_N\}$ randomly with probability $p_k$.
\item Do the corresponding random experiment with random variable $X_k$.
\end{itemize}
The final outcome can be described by an overall random variable $X$. Then it holds for $m\in\mathbb N$:
\begin{align}
\mathbb E[X^m] = \sum_{i=1}^N p_k\cdot \mathbb E[X_k^m]
\end{align}
\end{lemma}
\begin{proof}
Per definition for the expected value of $X^m$ (i.e. the $m$-th moment of $X$), we have to sum over the product of all possible outcomes and their corresponding probabilities. In our case, we can write that as
\begin{align}
\begin{split}
\mathbb E[X^m] &= \sum_{k=1}^N \sum_{\ell=1}^{n_k} x_{k\ell}^m \cdot p_k \cdot p_{k\ell} =  \sum_{k=1}^N p_k \cdot \underbrace{\sum_{\ell=1}^{n_k} x_{k\ell}^m  \cdot p_{k\ell}}_{=\mathbb E[X_k^m]} \\
& = \sum_{k=1}^N p_k \cdot \mathbb E[X_k^m],
\end{split}
\end{align}
which concludes the proof.
\end{proof}


 
 We now assume that the grain pattern of each granular medium has the same probability $1/N$. For the moments of the overall trajectory $V(x)$, we can apply Lemma \ref{lem::momente} for $m=1$ and $m=2$ to get
 \begin{align}
 \mathbb E[V(x)] &= \frac{1}{N}  \sum_{k=1}^N \mathbb E[V_k(x)],\label{eq::expectation}\\
 \mathbb E[V(x)^2] &= \frac{1}{N}  \sum_{k=1}^N \mathbb E[V_k(x)^2].
 \end{align}
 
 Those values can easily be determined by the results for each trajectory in Eq. \eqref{eq::1stmoment} and \eqref{eq::2ndmoment} and combination also leads to
 
 \begin{align}\label{eq::variance2}
 \mathbb V[V(x)] = \mathbb E[V(x)^2] - \mathbb E[V(x)]^2.
 \end{align}
 
 Finally with Eqs. \eqref{eq::expectation} and  \eqref{eq::variance2}, we have two formulas for the integrands in Eqs. \eqref{eq::sp} and \eqref{eq::np} and are able calculate the $\text{SP}$ and $\text{NP}$.\\
 In contrast to the magnetization mapping method of the previous section, we avoid the statistical error coming from the "projections" of the probabilities to the $-1$ and $+1$ magnetization states. The only statistical uncertainty probability mapping method comes from the random pattern of the grains in the different media.

 \section{Results and Discussion}
\begin{figure}[htbp]
\includegraphics[width=0.40\textwidth]{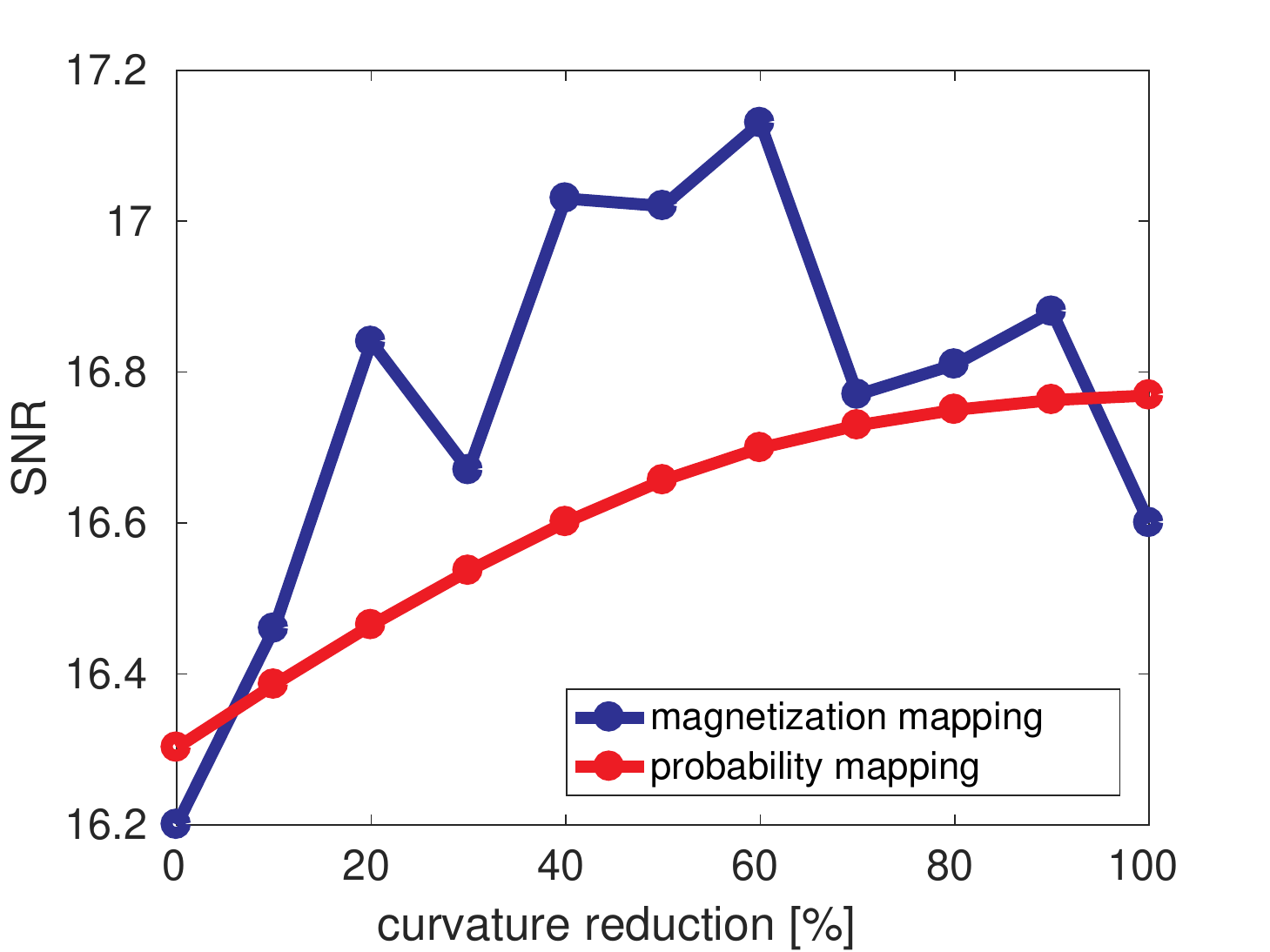}
\caption{Comparison of the SNR of a pseudo random bit sequence for varied bit curvature calculated by the methods of Secs. \ref{sec::writing} and \ref{sec::mapping}.}
\label{fig::compare}
\end{figure}

 We now compare the SNR calculation methods of Secs. \ref{sec::writing} (magnetization mapping) and \ref{sec::mapping} (probability mapping) based on a concrete example. Therefore we use the switching probability phase diagram in Fig. \ref{fig::phasenplot1} and for demonstration purpose we step-wise reduce the curvature of its shape (see Ref.~\onlinecite{b6}). For each curvature we calculate the SNR by writing on $N=50$ different granular media and determine the read-back signal $V(x)$ via a Gaussian shaped reader sensitivity function with a full width at half maximum of $13.26$~nm in down-track and $30.13$~nm in off-track direction. The SNR values are plotted in Fig. \ref{fig::compare}. We clearly observe that for the magnetization mapping method, the statistical variations on $50$ different granular media have large impact on the calculated SNR values, so no clear trend of the SNR can be observed. When we use the probability mapping method, we can massively reduce those variations and receive a much smoother curve with a clear trend, even tough the same number of granular media was used. The reason for the improvement is the avoided reduction of the magnetization probability to one of the magnetic end states $-1$ and $+1$ for each grain, which leads to a statistical error in the magnetization mapping method. Since each bit pattern contains a large number of grains, many repeated writings are necessary to obtain a sufficiently accurate mean read-back signal and its variance. The more sophisticated probability mapping writing method prevents this statistical issue and leads to very accurate results with comparatively few repetitions. Furthermore it is important to note that an implementation of both methods has asymptotically the same computational effort. In the writing process, the allocations of the magnetization or probabilities bases on similar approaches and have to be done $S$ times for $G$ grains in both methods. The additional calculation of $G$ times Eqs. \eqref{eq::e} and \eqref{eq::var} can be neglected. The read-back process requires summations over $R$ data points for every down-track position $x$ in both cases. The only noteworthy difference is that for the probability mapping method, it is unavoidable to calculate the grain sensitivity $H^i_G$ explicitly, since the square is used for the variance formula. This can be achieved by clever rearrangement of the sum and does not cause additional effort. In the final step the statistical evaluation in both cases is done via summation over $N$ granular media.

\section{Conclusion and Outlook}
In this work we presented a SNR calculation method based on probability theory. We could show that for a given switching probability phase diagram and given granular media it is statistically unfavorable to emulate the procedure of a real write processes in which each grain obtains a certain magnetic state according to its switching probability. Although it is the more obvious way to receive an estimate for the SNR, it reduces the switching probability $p\in[0,100]$\,\% to a binary value $m\in\{-1,+1\}$ representing the direction of the magnetization, which clearly means a loss of information. We were able to show that a prevention of this loss leads to a more sophisticated evaluation, which significantly improves the statistic and can furthermore be applied without additional computational effort. Although we demonstrated our approach as an application for SNR calculation in HAMR, similar ideas could also be used for conventional recording and other statistical evaluations on granular media. Since we successfully eliminated the statistical error of the reduction from $p$ to $m$, the only statistical error in our SNR value remains from the pattern of different granular media. The question arises whether this error could also be corrected by a suitable grain size distribution model instead of randomly sampling $N$ different media.

\section*{Acknowledgment}
The authors would like to thank the Austrian Science Fund (FWF) under grant No. I2214-N20 and the ASRC/IDEMA for financial support.

\end{document}